\documentclass[aps,pra,reprint,groupedaddress,showpacs]{revtex4-1}

\usepackage{graphicx}

\begin{document}

\title{Resonant self-pulsations in coupled nonlinear microcavities}

\author{Victor Grigoriev}
\email[]{victor.grigoriev@mpl.mpg.de}

\author{Fabio Biancalana}
\affiliation{Max Planck Institute for the Science of Light, G\"{u}nther-Scharowsky-Str. 1, Bau 26, Erlangen 91058, Germany}

\date{\today}

\begin{abstract}
A novel point of view on the phenomenon of self-pulsations is presented, which shows that they are a balanced state formed by two counteracting processes: beating of modes and bistable switching. A structure based on two coupled nonlinear microcavities provides a generic example of system with enhanced ability to this phenomenon. The specific design of such structure in the form of multilayered media is proposed, and the coupled mode theory is applied to describe its dynamical properties. It is emphasized that the frequency of self-pulsations is related to the frequency splitting between resonant modes and can be adjusted over a broad range.

\end{abstract}


\maketitle

\section{Introduction\label{sIntro}}

Optical bistability is a general nonlinear phenomenon, which can be observed in a large variety of systems provided that nonlinearity is complemented by some form of feedback. Apart from the classical Fabry-Perot resonators, the possible configurations of bistable systems include multilayered structures \cite{Grigoriev2010, Lidorikis2000}, fiber Bragg gratings \cite{Parini2007, Sterke1990} and microcavities in photonic crystal waveguides \cite{Maes2009, Soljacic2002}.

It was noticed however that the description of these systems in the frequency domain does not capture the full complexity of their temporal behavior. After overcoming the first switching threshold, the transient response of these systems may change dramatically, leading to self-pulsations or even chaos \cite{Ikeda1982, Ikeda1982a, Winful1982}. These instabilities in the time domain are usually explained from the viewpoint of nonlinear dynamics as a manifestation of the Hopf bifurcations \cite{Paulau2005, Hashemi2009, Tabor1989, Longhi2002, Trillo1996}. Although such description is appropriate, it is often desirable to have a control over these instabilities, as they might be useful for certain applications. In this work, we present a novel point of view on the phenomenon of self-pulsations which shows that they are a balanced state formed by two counteracting processes: beating of modes and bistable switching. We apply this concept to design a structure where self-pulsations can be excited very efficiently.

The paper is organized as follows. In Section \ref{sBeating}, we consider a typical example of the structure that supports self-pulsations. It is represented by the Bragg grating, and we explain how the beating of modes and bistable switching can interplay in this particular case. In Section \ref{sCoupled}, we propose a design of multilayered structure, which is based on two coupled nonlinear microcavities and shows pronounced self-pulsations. The structure has several independent parameters that make the adjustment of its properties easier. In Section \ref{sSwitching}, we derive a set of coupled mode equations which describes the switching dynamics in such system and discuss how the qualitative behavior of the structure agrees with the results obtained by the linear stability analysis. As a possible application, we show that self-pulsations can be used to convert a continuous wave signal into a regular train of ultrashort pulses. The last section summarizes the results and presents the conclusions.

\section{Beating of modes\label{sBeating}}

\begin{figure}
\includegraphics[width=80mm]{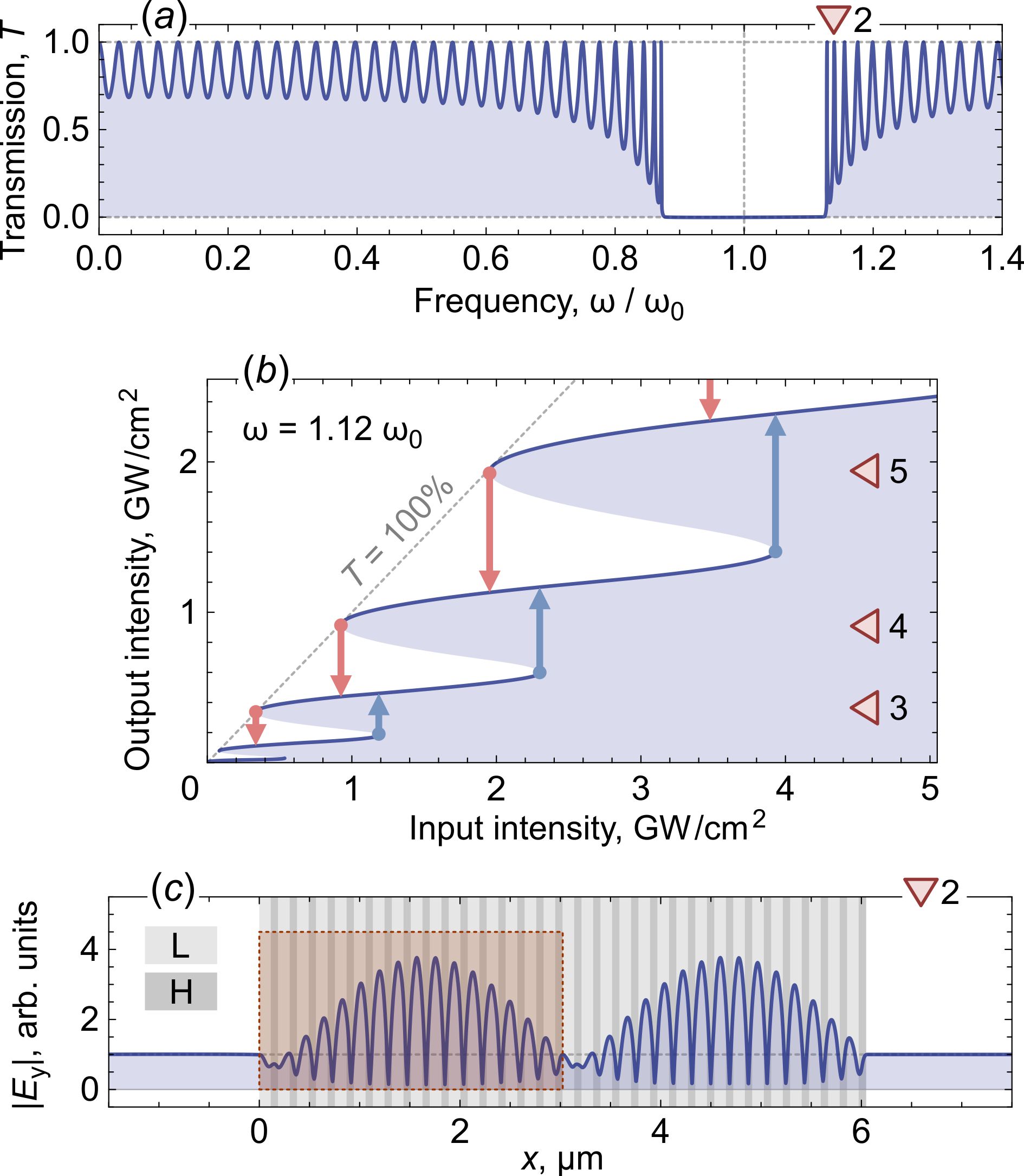}
\caption{\label{figBragg}(Color online).
(a)~Linear transmission spectrum of the Bragg structure consisting of 64 layers. For convenience, the resonances are counted from the upper band edge and marked by the red triangles with numbers. The positive Kerr nonlinearity can bend the linear resonances to the lower frequencies.
(b)~A typical hysteresis curve for signals with carrier frequency taken inside the band gap region. The loops created by the resonances for the increasing and decreasing input intensity are shown by arrows.
(c)~The linear field profile of the second resonance with one of the localization centers highlighted by the red rectangle. The background shows alternation of layers inside the structure.}
\end{figure}

As was mentioned in Introduction, self-pulsations tend to appear after the first switching threshold on the hysteresis curve. However, each loop of the hysteresis curve is related to a resonance in the linear transmission spectrum. For example, a Bragg structure obtained by periodic alternation of two different layers is considered on Fig.~\ref{figBragg}. It has many resonances near the band edge [Fig.~\ref{figBragg}(a)] which can be bent into the band gap region by the Kerr nonlinearity producing a multistable hysteresis curve [Fig.~\ref{figBragg}(b)]. Each resonance creates its own loop when input intensity is gradually increased and decreased. The difference between the first and higher switching thresholds is clearly pronounced in the field profiles of corresponding resonances. It turns out that the first resonance (counting from the upper band edge) has only a single localization center in the field profile, the second has two such centers [Fig.~\ref{figBragg}(c)], and this number continues to increase by one for the next resonances \cite{Lidorikis1996, Grigoriev2010a, Eilenberger2010}. Therefore, even the Bragg structure without any defects can be considered as a set of coupled microcavities.

It is known however that the beating of modes can exist in coupled systems. Although it is a purely linear effect, it can so strongly redistribute the energy between constituent parts of the system that they will meet conditions for up- and down-switching periodically. The need to overcome the first switching threshold in the case of Bragg structures is thus related to the fact that they can be approximated as a single-mode microcavity near the first resonance, and the beating of modes is not possible in such circumstances. When the input intensity reaches the switching threshold, the higher order resonances shifted by the Kerr nonlinearity will dominate, and the beating of modes can be excited. In principle, self-pulsations can be observed even without overcoming any threshold, if one chooses the carrier frequency to be initially near a higher order resonance.

There are two requirements for the onset of beating. Firstly, the system should support normal modes with slightly different frequencies. Most often, they can be built as symmetric and antisymmetric combinations of a degenerate mode. The difference of their frequencies determines the period of the beating. Secondly, the system should be excited in a way that creates an initial imbalance of energy between its parts, for example, by rapidly varying input intensity. The latter requirement is particularly important for open systems, because beating as any other transitional response is prone to decay in time. If the initial imbalance of energy is not regularly replenished, the system will settle in a stationary state. It turns out however that the systems of coupled nonlinear microcavities are able to sustain the beating of modes for arbitrary long time even if the input intensity is constant. This happens naturally because switching itself changes energy stored in microcavities in a step-like manner.

Therefore, switching and beating are closely related processes. Each of them can excite the other, and at the same time it needs to be sustained by the other. The interplay between these processes can take a periodic form, resulting in an infinite series of switchings, or self-pulsations.

\section{Coupled microcavities\label{sCoupled}}

The analysis performed in the previous section leads to conclusion that the simplest system with the ability to self-pulsations should consist of just two microcavities, and each of them can be strictly bistable. We consider a possible design of such system in the form of multilayered structure. It is sufficient to have dielectric layers of two types, which will be denoted as '$\rm{L}$' and '$\rm{H}$'. They correspond to materials with different refractive indices, and it will be assumed that the optical thickness of these layers satisfies the quarter wave condition. In what follows, the specific materials used for the layers '$\rm{L}$' and '$\rm{H}$' are polydiacetylene 9-BCMU with linear (nonlinear) refractive index  $n_{\rm{L}} = 1.55$ ($n_{2\rm{L}} = 2.5 \cdot 10^{-5} \; \rm{cm^2/MW}$) and rutile with $n_{\rm{H}} = 2.3$ ($n_{2\rm{H}} = 10^{-8} \; \rm{cm^2/MW}$), respectively~\cite{Tocci1995, Biancalana2008}. The quarter wave condition is set to $\lambda_0 = 0.7 \; \rm{\mu m}$, which gives the thicknesses of the layers $d_{\rm{L}} = 112 \; \rm{nm}$ and $d_{\rm{H}} = 76 \; \rm{nm}$.

\begin{figure}
\includegraphics[width=80mm]{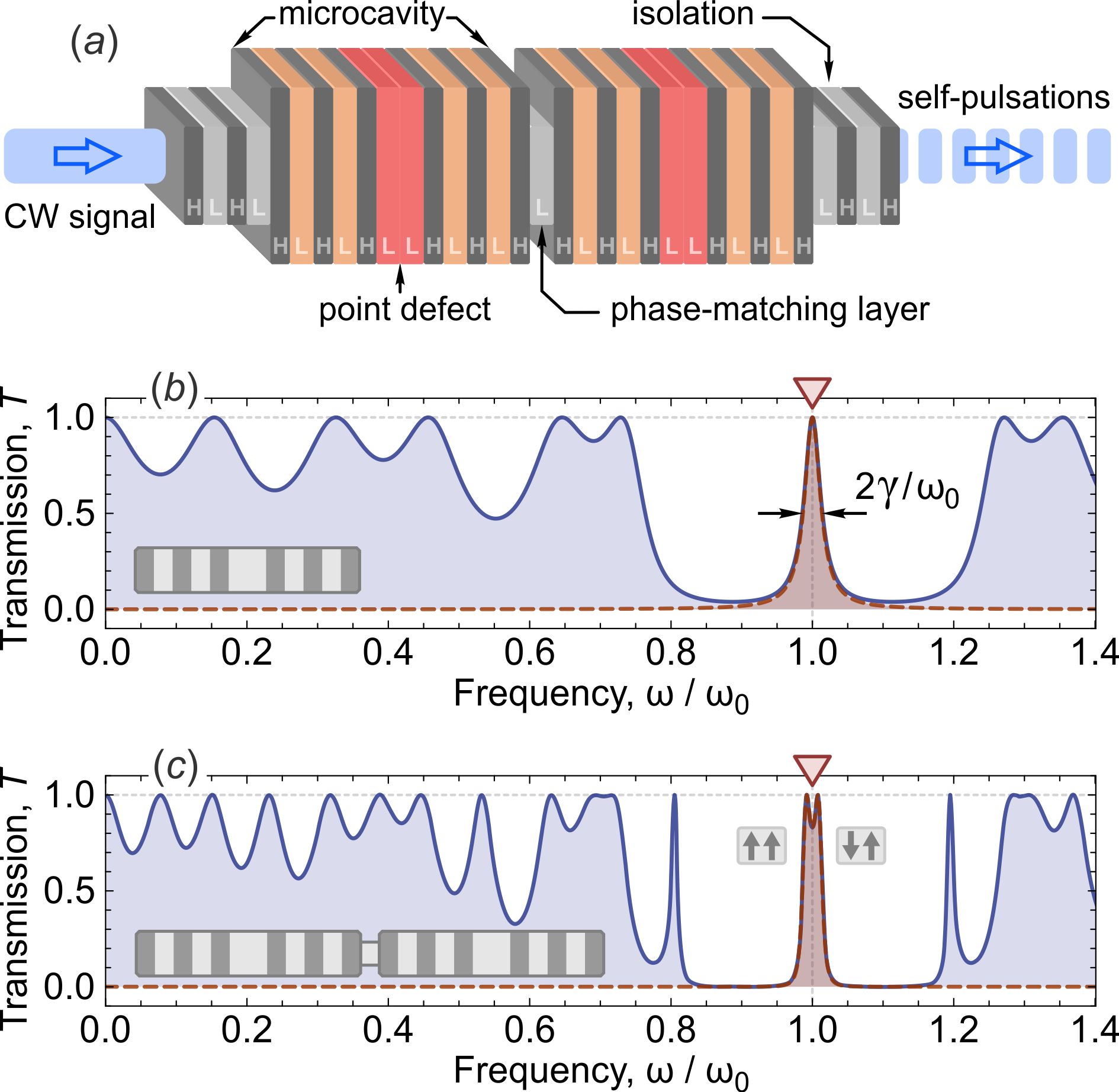}
\caption{\label{figStructure}(Color online).
(a)~Example of multilayered structure obtained from the symbolic formula (\ref{eqAB}), with various functional parts highlighted.
(b)~Transmission spectrum of the single microcavity for $m = 3$ (shown as an inset on the left) that was computed with the transfer matrix method (solid blue) and the coupled mode theory (dashed red). The resonance of perfect transmission caused by the point defect is in the middle of the band gap region, where the two spectra overlap.
(c)~Transmission spectrum of two coupled microcavities for $m = 3$ and $p = 0$. Due to the coupling, the degenerate resonance splits into a doublet of even and odd modes.}
\end{figure}

The microcavities can be formed by breaking periodicity in the arrangement of layers. We propose the following symbolic formula for the multilayered structure:
\begin{equation}
\label{eqAB}
(\textrm{HL})^{p + m} (\textrm{LH})^m
\textrm{L}
(\textrm{HL})^m (\textrm{LH})^{m + p},
\end{equation}
where the parameters $m$ and $p$ are integer numbers which satisfy the conditions $m \ge 1$ and $p \ge 0$. It is important to consider their meaning in more detail and to emphasize how they are related to various functional parts of the structure [Fig.~\ref{figStructure}(a)].

The two major parts are represented by the combination $(\textrm{HL})^m (\textrm{LH})^m$, which can be viewed as a result of imperfect junction of two periodic sequences. This causes the formation of a point-like defect in the middle, and creates a strong resonance in the transmission spectrum. This resonance is located exactly at the frequency for which the quarter wave condition holds, and the mirror symmetry of the structure ensures that it demonstrates perfect transmission [Fig.~\ref{figStructure}(b)]. The quality factor $Q$ of this resonance can be adjusted over a broad range by varying the number $m$. When the refractive indices of the layers $n_{\rm{L}}$ and $n_{\rm{H}}$ have a large contrast, the following formula can be used to estimate the quality factor
\begin{equation}
\label{eqQ}
Q = \frac {\omega_0} {2 \gamma} =  \frac{ \pi n_{\rm{H}} n_{\rm{L}} } { 4 (n_{\rm{H}}  - n_{\rm{L}} )}
    \left( \frac{ n_{\rm{H}} } { n_{\rm{L}} } \right) ^{2m}.
\end{equation}

The purpose of the single layer '$\textrm{L}$' placed in the middle of the structure is twofold. Firstly, it smoothly connects the two microcavities without introducing an additional defect into the structure. Secondly, it changes the phase mismatch between microcavities. If it were absent, the interaction between them would be suppressed, and the excitation of beating would not be possible. In addition, the structure can be extended with Bragg mirrors of the form $(\textrm{HL})^p$ and $(\textrm{LH})^p$ on the sides to make internal processes less affected by too early leakage of energy.

The influence of the coupling manifests itself already in the linear transmission spectrum, where the otherwise degenerate resonances of two identical microcavities split into a doublet [Fig.~\ref{figStructure}(c)]. Similar to the bonding and antibonding states of diatomic molecules, it is possible to distinguish the symmetric ($\omega < \omega_0$) and antisymmetric ($\omega > \omega_0$) modes in this doublet and to expect the onset of beating, when energy periodically tunnels from one microcavity to another. If the structure is nonlinear, these oscillations can modulate the optical properties leading to Raman-like scattering of incident waves on it and self-pulsations.

\section{Switching and self-pulsing\label{sSwitching}}

\begin{figure}
\includegraphics[width=80mm]{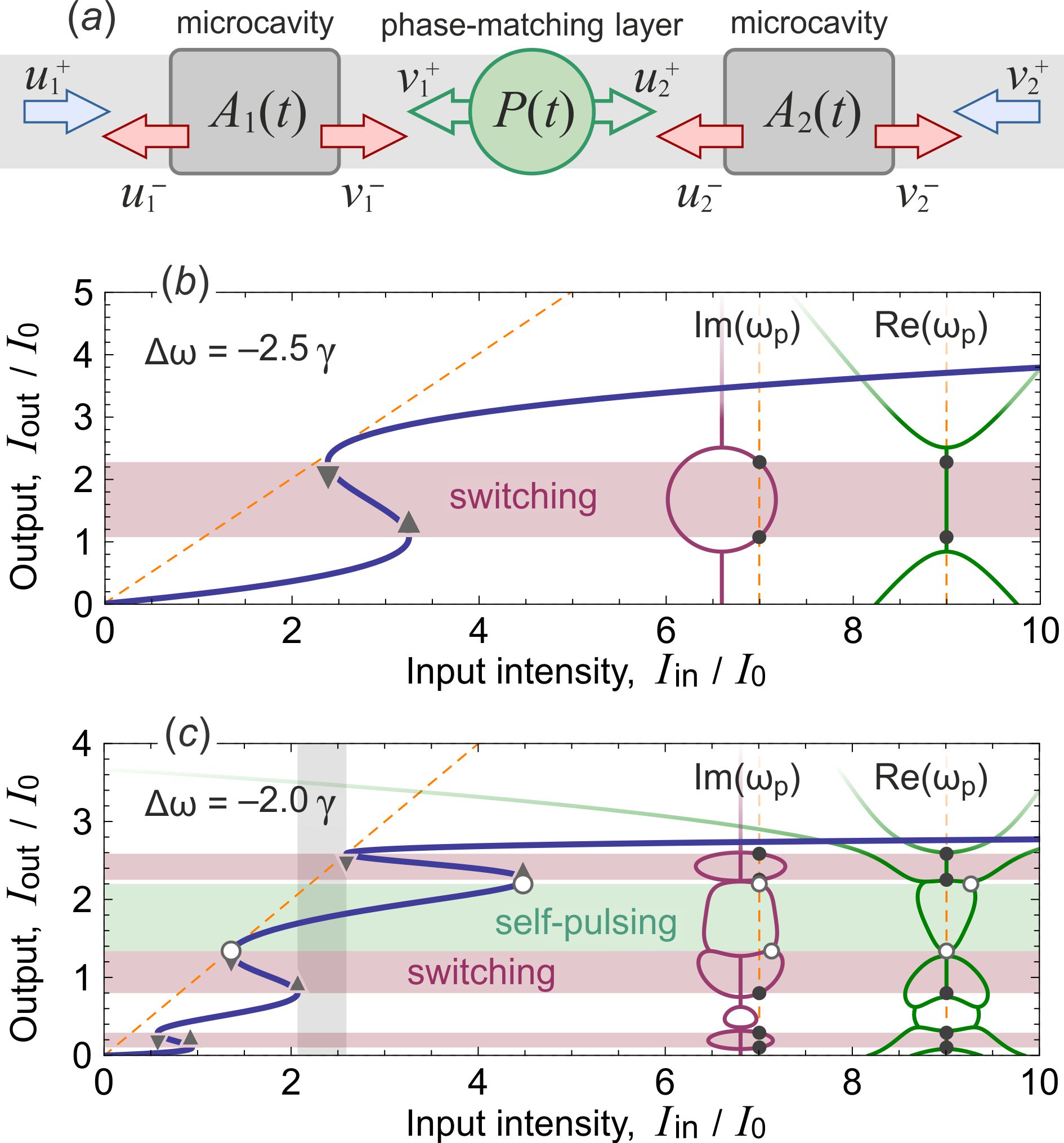}
\caption{\label{figModel}
(Color online).
(a)~A sketch of the model used to describe coupling between microcavities. The response of each element is approximated by a scattering matrix.
(b)~The hysteresis of output intensity for a single microcavity. The real and imaginary parts of eigenfrequencies obtained from the linear stability analysis are superimposed on the right side. Switching occurs when $\textrm{Im}(\omega_{\rm{p}}) > 0$ and $\textrm{Re}(\omega_{\rm{p}}) = 0$.
(c)~The hysteresis of output intensity for two coupled microcavities. Self-pulsing occurs when $\textrm{Im}(\omega_{\rm{p}}) > 0$ and $\textrm{Re}(\omega_{\rm{p}}) \neq 0$.
}
\end{figure}

To describe the switching dynamics in a single microcavity, we apply the coupled mode theory (CMT), which has been recognized as a useful tool for studying the properties of linear and nonlinear defects in photonic crystal waveguides \cite{Bravo-Abad2007, Fan2003, Haus1984}. Assuming that the microcavity is mirror symmetric, the CMT equations [see Fig.~\ref{figModel}(a) for notation] can be written as
\begin{equation}
\label{eqA}
\frac{dA}{dt} = \\
    - \left[ i  \left( \omega_0 - \gamma \frac{|A|^2}{I_0} \right) + \gamma \right]A \\
    + \gamma (u_+ + v_+ ),
\end{equation}
\begin{equation}
\label{eqUV}
\left( \begin{array}{c} u_- \\ v_- \\ \end{array} \right) \\
 =  - \left( \begin{array}{c} u_+ \\ v_+  \\ \end{array} \right) \\
 + A \left( \begin{array}{c} 1 \\ 1 \\ \end{array} \right),
\end{equation}
where $A(t)$ is the amplitude of the mode $E_0(x)$ with resonant frequency $\omega_0$ and damping constant $\gamma$, $u_\pm$($v_\pm$) are the amplitudes of in- and out-going plane waves on the left (right) side of the cavity.

Although the materials are assumed to be lossless, damping takes place due to the leakage of energy out of the structure and the decay rate can be estimated as
\begin{math}
\gamma  = [ (1 / c)
    \int_0^L{\varepsilon |E_0|^2 dx}
]^{-1}.
\end{math}
The integration in this formula is performed over the full length of the microcavity (spanning from 0 to $L$), $c$ is the speed of light in the vacuum, and  $\varepsilon(x)$ is the linear permittivity ($\varepsilon = n^2$). Since the electric field inside the microcavity is a product of the mode amplitude $A(t)$ and the mode profile $E_0(x)$, it is convenient to consider the mode profile as a dimensionless function which satisfies the boundary conditions $E_0(0) = E_0(L) = 1$. This ensures that the expression for the damping constant has a proper dimensionality.

The nonlinearity is responsible for the shift of resonant frequencies and can be taken into account by introducing the characteristic intensity for the microcavity
\begin{math}
I_0 = [ (\omega_0 / c)
    \int_0^L {n_2 \varepsilon |E_0| ^4 dx}
]^{-1}.
\end{math}
This formula follows from the orthogonal projection of the perturbation term which contains the nonlinear index of refraction $n_2(x)$ on the microcavity mode.

For the structure described by $m = 3$,  the damping constant and characteristic intensity take the following values $\gamma^{-1} = 27.6 \; \rm{fs}$,  $I_0 = 154 \; \rm{MW/cm^2}$. The question how accurate the CMT model is can be answered by comparing the linear transmission spectrum computed with a rigorous transfer matrix method and with the following formula derived from Eqs.~(\ref{eqA})--(\ref{eqUV})
\begin{equation}
\label{eqT}
T = \frac{ I_{\rm{out}} } { I_{\rm{in}} } = \left[ 1 + \left( \frac{ \omega  - \omega _0 }{ \gamma } +
\frac{ I_{\rm{out}} }{ I_{\rm{0}} } \right)^2 \right] ^ {-1},
\end{equation}
where $I_{\rm{in}} = |u_+|^2$ ($I_{\rm{out}} = |v_-|^2$) is the intensity of incident (transmitted) wave. Despite of the fact the quality factor of the considered resonance is rather low ($Q=37.2$), the two spectra overlap in a wide range of frequencies around $\omega_0$ [Fig.~\ref{figStructure}(b)].

For a fixed value of frequency detuning, Eq.~(\ref{eqT}) describes the hysteresis curve of the single microcavity [Fig.~\ref{figModel}(b)]. It is strictly bistable and the switching region can be determined from the linear stability analysis. The points on hysteresis curve, which have at least one eigenfrequency with positive imaginary part, cannot be stable, because small perturbations will exponentially grow forcing switching to another branch of hysteresis. It is worth noting that the eigenfrequencies of the linearized system $\omega_{\rm{p}}$ should satisfy a quadratic equation, $(\omega_{\rm{p}} / \gamma + i)^2 + q = 0$, where $q$ is a real coefficient. Therefore, the transient response after the switching should have the form of exponentially decaying oscillations $\textrm{exp}(-i \omega_{\rm{p}} t)$, where $\omega_{\rm{p}} / \gamma = \pm \sqrt{|q|} - i$.

A qualitatively different behavior is possible in presence of two coupled microcavities. The corresponding CMT equations can be written as
\begin{equation}
\label{eqA1}
\frac{dA_1}{dt} = \\
    - \left[ i \left( \omega_0 - \gamma \frac{|A_1|^2}{I_0} \right) + \frac{\gamma}{2 \kappa^2} \right]A_1 \\
    + i \frac{n_{\rm{L}}\gamma}{2} A_2 + \frac{\gamma}{\kappa} u_1^+,
\end{equation}
\begin{equation}
\label{eqA2}
\frac{dA_2}{dt} = \\
    - \left[ i \left( \omega_0 - \gamma \frac{|A_2|^2}{I_0} \right) + \frac{\gamma}{2 \kappa^2} \right]A_2 \\
    + i \frac{n_{\rm{L}}\gamma}{2} A_1 + \frac{\gamma}{\kappa} v_2^+,
\end{equation}
where $n_{\rm{L}}$ is the refractive index of the phase-matching layer '$\textrm{L}$', and the parameter $\kappa = (-  n_{\rm{H}} / n_{\rm{L}} )^p$ takes into account the influence of the Bragg mirrors. The linear normal modes of the system~(\ref{eqA1})--(\ref{eqA2}) are given by combinations $A_1 \pm A_2$, and their resonant frequencies are $\omega = \omega_0 \mp n_{\rm{L}} \gamma / 2$. It is important that the frequency splitting depends only on the number $m$ and can be estimated using Eq.~(\ref{eqQ}), while the number $p$ determines the external quality factor $Q_{\rm{ext}} = \kappa^2 Q$  and can lower the effective characteristic intensity  $I_{\rm{eff}} = I_0 / \kappa^4$. The latter property can be useful if materials have a weak Kerr nonlinearity, but for simplicity it will be assumed that $p = 0$  and consequently $\kappa = 1$.

\begin{figure}
\includegraphics[width=80mm]{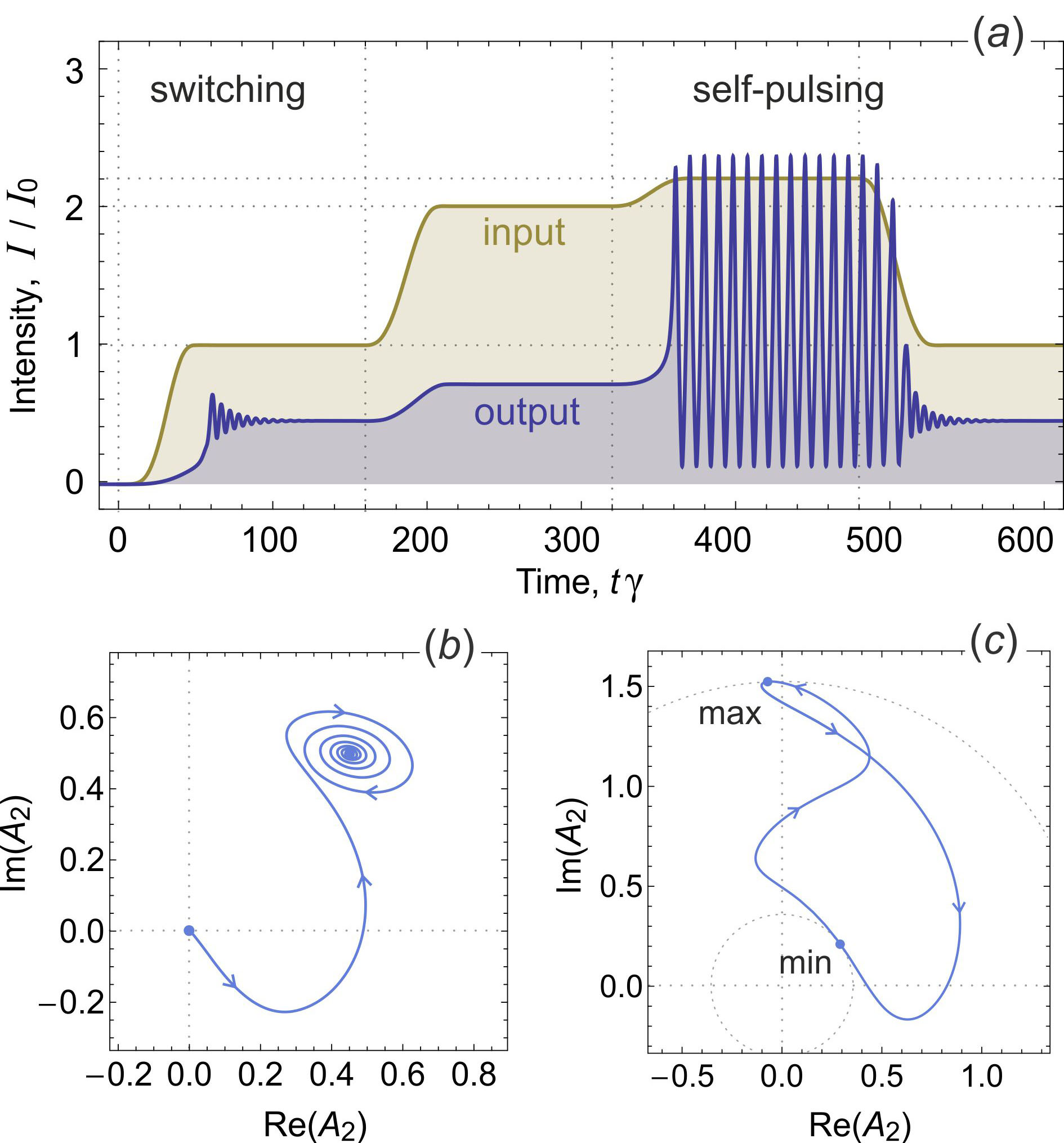}
\caption{\label{figDynamics}
(Color online).
(a)~The intensities of incident and transmitted signals as functions of time during the switching between stable states and after transition to the self-pulsations.
(b)~A phase portrait of the switching, which is obtained from the real and imaginary parts of transmission amplitude. The final state corresponds to a stable spiral point.
(c)~A similar phase portrait for one period of self-pulsations. A limit cycle is formed due to the Hopf bifurcation~\cite{Tabor1989}.
}
\end{figure}

The hysteresis curve of this system shows that there is a range of input intensities where only unstable solutions are possible [vertical gray region in Fig.~\ref{figModel}(c)].  This range falls exactly between the resonances of the doublet, and the transient dynamics in this region is different from the switching [see Fig.~\ref{figDynamics} and caption for details]. In the time domain simulation, the input intensity of a continuous wave signal was slowly varied to examine stability of several key points on the hysteresis curve. After overcoming the second threshold the system goes into a state of infinite switching. The period of these pulses extracted from the simulation data was 252~fs, which is of the same order of magnitude as the beating period of the two resonant modes $4 \pi / (n_{\rm{L}} \gamma) = 224\; \rm{fs}$. Therefore, the frequency of self-pulsations $\omega_{\rm{sp}} = 4 \; \rm{THz}$ can be estimated by the following formula
\begin{equation}
\label{eqWsp}
\omega_{\rm{sp}} = (n_{\rm{L}} \omega_0) / (4 Q).
\end{equation}
It shows explicitly that smaller internal quality factors can produce train of pulses with higher repetition rate, since photons tend to spend less time in each cavity and they exchange energy at a faster rate. This property can be useful in optical communications, where such pulses can transfer bits of information~\cite{Coen2001}. Another possible application is optical clocks for photonic circuits, where train of pulses can synchronize the operation of different components \cite{Papakyriakopoulos1999}.

There are two reasons why the frequency of beating and self-pulsations do not coincide precisely. Firstly, the structure considered in this paper belongs to the open systems, and thus the resonant frequencies are not real quantities. Depending on the value of the quality factor, this can limit accuracy to about 1\%. Secondly, the Kerr nonlinearity changes the refractive indices of the layers and consequently shifts all resonant frequencies. This can increase the discrepancy even further, but since the period of the self-pulsations is determined by the frequency splitting of the doublet rather than its absolute position, the accuracy of about 10\% can be considered as acceptable to prove the concept.

The quantitative agreement can be improved if self-pulsations are not very intense. As an example we can apply our interpretation of self-pulsations to a paper of other authors \cite{Maes2009}, where this phenomenon was studied for microcavities embedded in photonic crystal waveguides. We derived an explicit formula for the frequency of beating in their case and found that it differs from the frequency of self-pulsations only in 4\%. It is worth noting that the concept of gap solitons \cite{Sterke1990, Lidorikis2000} which is often used as a physical explanation of self-pulsations does not give any estimations at all in this case.

\section{Conclusions\label{sConclusions}}

In conclusion, self-pulsations can be viewed as a balanced state formed by switching and beating of modes in coupled nonlinear microcavities. We proposed the design of multilayered structure where self-pulsations can be excited very efficiently. Since the phenomenological CMT equations were applied to describe the switching dynamics, the specific choice of configuration is not critical, and we expect qualitatively similar results in other configurations and physical systems.

It is also possible to observe self-pulsations in systems with continuous spectrum such as waveguide couplers or dual-core fibers~\cite{Daino1985, Trillo1989}. The coupling between waveguides leads to the beating of modes in space, and the overall dynamics can be described by equations which are similar to Eqs.~(\ref{eqA1})--(\ref{eqA2}). However, it is not easy to excite selectively odd or even modes in that case~\cite{Agrawal2008}, and therefore the usage of coupled microcavities offers some advantage.

\begin{acknowledgments}
This work was supported by the German Max Planck Society for the Advancement of Science (MPG).
\end{acknowledgments}

\bibliography{references}

\end{document}